\documentclass{article}

\usepackage{PRIMEarxiv}

\usepackage[utf8]{inputenc} 
\usepackage[T1]{fontenc}    
\usepackage{hyperref}       
\usepackage{url}            
\usepackage{booktabs}       
\usepackage{amsfonts}    
\usepackage{amsmath}
\usepackage{amssymb}
\usepackage{nicefrac}       
\usepackage{microtype}      
\usepackage{lipsum}
\usepackage{fancyhdr}       
\usepackage{graphicx}       
\graphicspath{{media/}}     
\usepackage{braket}
\usepackage{mathtools}
\usepackage{hhline}
\usepackage{tikz}
\usetikzlibrary{calc}
\usepackage{authblk}
\usepackage{subcaption}
\usepackage{amsthm}
\usepackage{algorithm}
\usepackage{algpseudocode}

\usepackage{changes}

\renewcommand{\ket}[1]{{| #1 \rangle}}

\newtheorem*{conjecture}{Conjecture}

\newtheorem{proposition}{Proposition}

\title{A counterexample to the Ho–Hsieh \\VQCS saturation conjecture}

\author[1]{Edward Jiang\thanks{ye-duo.jiang@icfo.eu}\,\,\,}

\affil[1]{ICFO -- Institut de Ciencies Fotoniques, The Barcelona Institute of Science and Technology,\newline 08860 Castelldefels, Barcelona, Spain}

\begin{document}
\maketitle
\begin{abstract}
We show that there exists a 4-qubit state that can be prepared by a variational quantum-classical simulation (VQCS) circuit ansatz containing 3 layers, but not with 2 layers. This disproves the strong form of a conjecture presented by Ho and Hsieh stating that the expressivity of VQCS on $L$ qubits saturates at $L/2$ layers. The weak form of the conjecture concerning the preparation of the ground state remains open.
\end{abstract}

\section{Introduction}

The variational quantum-classical simulation (VQCS) ansatz is a variational ansatz originally used to prepare ground states of the transverse-field Ising model (TFIM), inspired by the quantum approximate optimization algorithm (QAOA) \cite{farhi2014quantum,farhi2016quantum} and variational quantum eigensolvers (VQE) \cite{peruzzo2014variational,wecker2015progress}. The ansatz consists of alternating layers between the two Hamiltonians
\begin{align}
H_X&=-\sum_{i=1}^NX_i\qquad\qquad\qquad
H_{ZZ}=-\sum_{i=1}^N Z_iZ_{i+1}
\end{align}
so that, for a circuit with $p$ layers, parameter vectors $\vec\gamma=(\gamma_1,\cdots,\gamma_p)$ and $\vec\beta=(\beta_1,\cdots,\beta_p)$, and initial state $\ket{\Psi_0}=\ket{{+}{+}\cdots{+}}$, the final state is given by
\begin{align}
\ket{\Psi;p,\vec\gamma,\vec\beta}=\left(\prod_{k=1}^p\exp\left(-i\beta_kH_X\right)\exp\left(-i\gamma_kH_{ZZ}\right)\right)\ket{\Psi_0}
\end{align}
The primary aim is to variationally prepare ground states of the TFIM with periodic boundaries, which has the Hamiltonian $H_{\text{TFIM}}=H_{ZZ}+gH_X$ for any chosen $g\in\mathbb{R}$. 

The authors of \cite{ho2019efficient} presented the following conjectures on the saturation of the expressivity of VQCS:
\begin{conjecture}[Strong form]
For even $L$, any state that can be prepared by a VQCS circuit with $p\geq 1$ layers and parameters $(\vec{\gamma}, \vec\beta)$ can also be prepared by a VQCS circuit with $L/2$ layers and other parameters $(\vec\gamma ',\vec\beta')$.
\end{conjecture}
\begin{conjecture}[Weak form]
For even $L$ and any field strength $g$, the ground state of the TFIM can be prepared by a VQCS circuit with $L/2$ layers.
\end{conjecture}
The strong form is sufficient but not necessary for the weak form of the conjecture to hold. Indeed, the weak form follows trivially from the strong form by Trotterizing the evolution of a quantum adiabatic algorithm \cite{farhi2001quantum,farhi2000quantum}, which prepares the exact ground state as $p\to\infty$. Evidence for the weak form has been established in the literature both analytically using the free-fermion solution \cite{dreyer2021quantum} and also numerically \cite{arezzo2026digital,matos2023characterization}. More recently, techniques from quantum signal processing were used to prove the related Farhi-Goldstone-Gutmann conjecture \cite{farhi2014quantum} concerning the case where $p<L/2$ \cite{kol2026machine,marwaha2026qaoa}.

\section{Main result}

We show that the strong form of the conjecture is false by providing a counterexample in the $L=4$ case. First define the 4-qubit states
\begin{align}
\ket{\Psi_R}&=\frac{1}{2}\left(\ket{{+}{+}{-}{-}}+\ket{{-}{+}{+}{-}}+\ket{{-}{-}{+}{+}}+\ket{{+}{-}{-}{+}}\right)\\
\ket{\Psi_S}&=\frac{1}{\sqrt{2}}\left(\ket{{+}{-}{+}{-}}+\ket{{-}{+}{-}{+}}\right)
\end{align}
as well as the target state
\begin{align}
\ket{\Psi_T}=\frac{1}{\sqrt{2}}\left(\ket{\Psi_R}+\ket{\Psi_S}\right)
\end{align}

\begin{proposition}
$\ket{\Psi_T}$ cannot be prepared by a VQCS circuit with depth $p=2$.
\end{proposition}
\begin{proof}
We move to the free-fermion pseudospin picture \cite{mbeng2020quantum,wang2018quantum} via an isometry $V$ of the 4-dimensional subspace spanned by $\left\{\ket{{+}{+}{+}{+}},\ket{{-}{-}{-}{-}},\ket{\Psi_R},\ket{\Psi_S}\right\}$:
\begin{align}
V:\qquad\  
\begin{aligned}
\ket{{+}{+}{+}{+}}&\quad\mapsto\quad \ket{{+}{+}}\\
\ket{{-}{-}{-}{-}}&\quad\mapsto\quad \ket{{-}{-}}\\
\ket{\Psi_R}&\quad\mapsto\quad \tfrac{1}{\sqrt{2}}\left(\ket{{+}{-}}+\ket{{-}{+}}\right)\\
\ket{\Psi_S}&\quad\mapsto\quad \tfrac{1}{\sqrt{2}}\left(\ket{{+}{-}}-\ket{{-}{+}}\right)
\end{aligned}
\end{align}
In this picture, the Hamiltonians $H_X$ and $H_{ZZ}$ transform as
\begin{align}
\tilde{H}_X&=VH_XV^\dagger=-2\left(X\otimes I+I\otimes X\right)\\
\tilde{H}_{ZZ}&=VH_{ZZ}V^\dagger=-2\left(\left(\frac{Z+X}{\sqrt{2}}\right)\otimes I+I\otimes \left(\frac{Z-X}{\sqrt{2}}\right)\right)
\end{align}
and the initial and target states become
\begin{align}
\ket{\tilde{\Psi}_0}=V\ket{\Psi_0}&=\ket{{+}{+}}\qquad\qquad\qquad \ket{\tilde{\Psi}_T}=V\ket{\Psi_T}=\ket{{+}{-}}
\end{align}
Towards a contradiction, suppose that there exist some $\gamma_1,\beta_1,\gamma_2,\beta_2$ such that
\begin{align}
\ket{\tilde{\Psi}_T}=\exp\left(-i\tilde{H}_X\beta_2\right)\exp\left(-i\tilde{H}_{ZZ}\gamma_2\right)\exp\left(-i\tilde{H}_X\beta_1\right)\exp\left(-i\tilde{H}_{ZZ}\gamma_1\right)\ket{\tilde{\Psi}_0}
\end{align}
Each of the two decoupled pseudospins gives one condition:
\begin{align}
\ket{{+}}&=\exp\left(2i\beta_2X\right)\exp\left(2i\gamma_2\frac{Z+X}{\sqrt{2}}\right)\exp\left(2i\beta_1X\right)\exp\left(2i\gamma_1\frac{Z+X}{\sqrt{2}}\right)\ket{{+}}\\
\ket{{-}}&=\exp\left(2i\beta_2X\right)\exp\left(2i\gamma_2\frac{Z-X}{\sqrt{2}}\right)\exp\left(2i\beta_1X\right)\exp\left(2i\gamma_1\frac{Z-X}{\sqrt{2}}\right)\ket{{+}}
\end{align}
Taking inner products with $\ket{{-}}$ and $\ket{{+}}$ respectively, these conditions become
\begin{align}
\exp(2i\beta_1)\sin2\gamma_2\left(\cos2\gamma_1+\frac{i}{\sqrt{2}}\sin2\gamma_1\right)+\exp(-2i\beta_1)\sin2\gamma_1\left(\cos2\gamma_2-\frac{i}{\sqrt{2}}\sin2\gamma_2\right)&=0\label{eq:cond1}\\
\exp(2i\beta_1)\left(\cos2\gamma_1-\frac{i}{\sqrt{2}}\sin2\gamma_1\right)\left(\cos2\gamma_2-\frac{i}{\sqrt{2}}\sin2\gamma_2\right)-\frac{1}{2}\exp(-2i\beta_1)\sin2\gamma_1\sin2\gamma_2&=0\label{eq:cond2}
\end{align}
Now, taking the norm squared of the two terms in (\ref{eq:cond2}) forces $\sin^22\gamma_1+\sin^22\gamma_2=2$, so $\sin^22\gamma_1=\sin^22\gamma_2=1$ and $\cos2\gamma_1=\cos2\gamma_2=0$. Then (\ref{eq:cond1}) implies that $\exp(2i\beta_1)-\exp(-2i\beta_1)=0$, equivalently $\exp(4i\beta_1)=1$. On the other hand, (\ref{eq:cond2}) implies that $\exp(2i\beta_1)+\exp(-2i\beta_1)=0$, equivalently $\exp(4i\beta_1)=-1$.
\end{proof}

\begin{proposition}
$\ket{\Psi_T}$ can be prepared by a VQCS circuit with depth $p=3$.
\end{proposition}
\begin{proof}
Take $\vec{\gamma} = (\pi/4, 5\pi/16, 3\pi/8)$ and $\vec{\beta}= (3\pi /8, \pi/4, 0)$.
\end{proof}

\section{Conclusion}

We provided a counterexample to the strong form of the conjecture in the case where $L=4$. Not discussed is whether the argument can be generalized to $L\geq 6$, though this may be possible by introducing recent techniques from \cite{kol2026machine,marwaha2026qaoa}. Also not discussed is whether the weak form of the conjecture holds.

\bibliographystyle{plain}
\bibliography{references}

@article{ho2019efficient,
  title={Efficient variational simulation of non-trivial quantum states},
  author={Ho, Wen Wei and Hsieh, Timothy H},
  journal={SciPost Physics},
  volume={6},
  number={3},
  pages={029},
  year={2019}
}

@article{marwaha2026qaoa,
  title={The {{QAOA}} on the ring of disagrees},
  author={Marwaha, Kunal},
  journal={arXiv preprint arXiv:2606.29562},
  year={2026}
}

@article{kol2026machine,
  title={A Machine-Verified Proof of a Quantum-Optimization Conjecture},
  author={Kol, Uri and Ben-Shahar, Maor and Sulimany, Kfir and Englund, Dirk},
  journal={arXiv preprint arXiv:2606.29687},
  year={2026}
}

@article{farhi2014quantum,
  title={A quantum approximate optimization algorithm},
  author={Farhi, Edward and Goldstone, Jeffrey and Gutmann, Sam},
  journal={arXiv preprint arXiv:1411.4028},
  year={2014}
}

@article{farhi2016quantum,
  title={Quantum supremacy through the quantum approximate optimization algorithm},
  author={Farhi, Edward and Harrow, Aram W},
  journal={arXiv preprint arXiv:1602.07674},
  year={2016}
}

@article{wecker2015progress,
  title={Progress towards practical quantum variational algorithms},
  author={Wecker, Dave and Hastings, Matthew B and Troyer, Matthias},
  journal={Physical Review A},
  volume={92},
  number={4},
  pages={042303},
  year={2015},
  publisher={APS}
}

@article{peruzzo2014variational,
  title={A variational eigenvalue solver on a photonic quantum processor},
  author={Peruzzo, Alberto and McClean, Jarrod and Shadbolt, Peter and Yung, Man-Hong and Zhou, Xiao-Qi and Love, Peter J and Aspuru-Guzik, Al{\'a}n and O’brien, Jeremy L},
  journal={Nature communications},
  volume={5},
  number={1},
  pages={4213},
  year={2014},
  publisher={Nature Publishing Group UK, London}
}

@article{mbeng2020quantum,
  title={The quantum {{Ising}} chain for beginners},
  author={Mbeng, Glen Bigan and Russomanno, Angelo and Santoro, Giuseppe E},
  journal={networks},
  volume={16},
  pages={17},
  year={2020}
}

@article{wang2018quantum,
  title={Quantum approximate optimization algorithm for MaxCut: A fermionic view},
  author={Wang, Zhihui and Hadfield, Stuart and Jiang, Zhang and Rieffel, Eleanor G},
  journal={Physical Review A},
  volume={97},
  number={2},
  pages={022304},
  year={2018},
  publisher={APS}
}

@article{farhi2000quantum,
  title={Quantum computation by adiabatic evolution},
  author={Farhi, Edward and Goldstone, Jeffrey and Gutmann, Sam and Sipser, Michael},
  journal={arXiv preprint quant-ph/0001106},
  year={2000}
}

@article{farhi2001quantum,
  title={A quantum adiabatic evolution algorithm applied to random instances of an NP-complete problem},
  author={Farhi, Edward and Goldstone, Jeffrey and Gutmann, Sam and Lapan, Joshua and Lundgren, Andrew and Preda, Daniel},
  journal={Science},
  volume={292},
  number={5516},
  pages={472--475},
  year={2001},
  publisher={American Association for the Advancement of Science}
}

@article{dreyer2021quantum,
  title={Quantum computing critical exponents},
  author={Dreyer, Henrik and Bejan, Mircea and Granet, Etienne},
  journal={Physical Review A},
  volume={104},
  number={6},
  pages={062614},
  year={2021},
  publisher={APS}
}

@article{matos2023characterization,
  title={Characterization of variational quantum algorithms using free fermions},
  author={Matos, Gabriel and Self, Chris N and Papi{\'c}, Zlatko and Meichanetzidis, Konstantinos and Dreyer, Henrik},
  journal={Quantum},
  volume={7},
  pages={966},
  year={2023},
  publisher={Verein zur F{\"o}rderung des Open Access Publizierens in den Quantenwissenschaften}
}

@article{arezzo2026digital,
  title={Digital controllability of transverse-field Ising chains},
  author={Arezzo, Vincenzo Roberto and Wang, Ruiyi and Thengil, Kiran and Pecci, Giovanni and Santoro, Giuseppe E},
  journal={Physical Review A},
  volume={113},
  number={1},
  pages={012610},
  year={2026},
  publisher={APS}
}

\end{document}